%% file: paper.tex
\documentclass[%
twocolumn,
floatfix,
 reprint,
 amsmath,amssymb,
 aps,
]{revtex4-2}

\usepackage{graphicx}% Include figure files
\usepackage{dcolumn}% Align table columns on decimal point
\usepackage{bm}% bold math
\usepackage{enumerate}
\usepackage{comment}
\usepackage{hyperref}% add hypertext capabilities
\usepackage{float}

\preprint{APS/123-QED}
\input{def.tex}

\begin{document}

\preprint{APS/123-QED}

\title{Exploring hadronic rescattering effects on resonance productions \\ in pp and p--Pb collisions with PYTHIA8}% Force line breaks with \\
\author{Su-Jeong Ji} \email{su-jeong.ji@cern.ch} \affiliation{\pusan}
\author{Gyeongbin Park} \affiliation{\pusan} 
\author{Bong-Hwi Lim}  \affiliation{\infn}
\author{Sanghoon Lim} \email{shlim@pusan.ac.kr} \affiliation{\pusan}\affiliation{\extreme}

\newcommand{\infn}{INFN Sezione di Torino, Turin 10125, Italy}
\newcommand{\pusan}{Department of Physics, Pusan National University, Busan 46241, South Korea}
\newcommand{\extreme}{Extreme Physics Institute, Pusan National University, Busan 46241, South Korea}

\date{\today}% It is always \today, today,
             %  but any date may be explicitly specified2

\begin{abstract}
%% ADD HADONRIC PHASE IN HEAVY-ION COLLISIONS %%
In relativistic heavy-ion collisions, the quark-gluon plasma is created, and as the medium cools down, the system transitions into a hadronic phase.
While such interactions are well established for large systems, such as Pb--Pb collisions, their relevance in smaller collision systems remains unclear.
Consequently, hadronic interactions during the hadronic phase are studied in pp collisions at $\sqrt{s}=13$ TeV and p--Pb collisions at $\sqrt{s_{\rm{NN}}}=5.02$ TeV with the PYTHIA8 event generator.
The interaction is studied via the yield ratios between resonances and stable particles with similar quark contents, which are obtained as a function of transverse momentum ($p_{\rm{T}}$) using $\mathrm{\rho(770)^0}$, $\mathrm{K^*(892)^0}$, and $\mathrm{\phi(1020)}$ mesons and their stable particles, $\mathrm{\pi^\pm}$ and $\mathrm{K^\pm}$ at midrapidity ($|\rm{y}|<0.5$).
Yield ratios are calculated in five multiplicity classes for pp and six for p--Pb collisions, using the 60--100\% multiplicity class in pp as a reference.
Although rescattering leads to stronger suppression at low $p_{\rm{T}} < 2$ GeV/$c$, a visible suppression remains even when rescattering is turned off.
To isolate the rescattering effect, double ratios between the rescattering on and off configurations are obtained.
These are then integrated in the full \pt\ range ($0<p_{\rm{T}}<6.0$ GeV/$c$).
The normalized double ratios show a decreasing trend with increasing multiplicity, independent of the collision system.
The lower limit of the hadronic phase lifetimes extracted in integrated-\pt\ region increases with multiplicity in both systems, but with a notable discrepancy between pp and p--Pb collisions.

%The ratios are calculated in five multiplicity classes for pp and six multiplicity classes for p--Pb collisions, referencing the 60--100\% from pp collisions.
%Even though the yield ratios with rescattering show larger suppression for $p_{\rm{T}} < 2$ GeV/$c$, suppression is still observed for the rescattering off option.
%To consider the effects of rescattering alone, double ratios between the yield ratios of the rescattering on and rescattering off options are obtained.
%And then the double ratios to pp 60--100\% in $0<p_{\rm{T}}<6.0$ GeV/$c$, which is the total $p_{\rm T}$ range and in $0<p_{\rm{T}}<0.8$ GeV/$c$, where the rescattering effect is maximized, are integrated as a function of multiplicity in both collision systems to see the system size dependent.
%The normalised double ratios decrease to 60--100\% with increasing multiplicity, showing a similar trend regardless of the system size.
%The lifetime of the hadronic phase is also calculated in pp and p--Pb collisions in $0<p_{\rm{T}}<0.8$ GeV/$c$, and the results show an increasing trend with increasing multiplicity, as expected. However, the values exhibit a discrepancy between pp and p--Pb collisions. 
\end{abstract}

%\keywords{Suggested keywords}%Use showkeys class option if keyword
                              %display desired
\maketitle

\section{\label{sec:intro}Introduction}

%At relativistic heavy-ion collisions such as the energy obtained at the Relativistic Heavy Ion Collider (RHIC) and the Large Hadron Collider (LHC), quark-gluon plasma (QGP) is thought to be created~\cite{PHENIX:2004vcz, STAR:2005gfr, PHOBOS:2004zne, BRAHMS:2004adc, ALICE:2022wpn}. 
%As QGP expands, it cools down and hadrons are formed at chemical freeze-out. In this hadronic gas phase, hadrons still interact with each other until the system reaches kinetic freeze-out. 
%This hadronic phase lasts for over tens of fm/$c$, which is compatible with the lifetime of several resonances. 
%Therefore, some of the hadrons with short lifetimes, such as resonances, decay during the hadronic phase. 
%The decay products of these resonances can undergo interactions in the hadronic phase with nearby hadrons, changing their momenta, making it difficult to reconstruct the original mother particles. 
%The degree of these interactions, such as rescattering and regeneration, can differ depending on the lifetime of the hadronic phase, which is known to be longer in larger collision systems. 
%Therefore, resonances can be a suitable probe for investigating the properties of the hadronic phase. %~\cite{ALICE:2019xyr}.

In high-energy heavy-ion collisions, such as those conducted at the Relativistic Heavy Ion Collider (RHIC) and the Large Hadron Collider (LHC), a deconfined state of quarks and gluons—known as the quark-gluon plasma (QGP)—is believed to be formed in the early stages of the collision~\cite{PHENIX:2004vcz, STAR:2005gfr, PHOBOS:2004zne, BRAHMS:2004adc, ALICE:2022wpn}. 
As the QGP expands and cools, the system undergoes a phase transition to a hadron gas at chemical freeze-out, where hadron species are fixed. 
However, these hadrons continue to interact with one another until the system reaches kinetic freeze-out~\cite{Rapp:2000gy, Song:1996ik, Rafelski:2001hp}. 
This intervening hadronic phase can last for several tens of femtometers per $c$ (fm/$c$), comparable to the lifetimes of many short-lived resonances~\cite{ParticleDataGroup:2024cfk}. 
As a result, resonances that decay during this stage may have their decay products modified by subsequent interactions with other hadrons in the medium. 
Such interactions—predominantly rescattering, where the decay products undergo elastic or inelastic collisions that distort their original momentum correlation, and regeneration, where a resonance is reformed from two hadrons in the medium—can significantly alter the observed resonance yields and momentum distributions.
The strength and frequency of these hadronic interactions depend on the duration and density of the hadronic phase, which are typically longer and higher in larger collision systems.
%SUch interactions-predominantly rescattering and regeneration-can alter the momenta of the decay products, complication the reconstruction of the original resonance signals.
%Significantly, the strength and frequency of these hadronic interactions depend on the duration and density of the hadronic phase, which are typically longer and higher in larger collision systems.
Therefore, resonances serve as sensitive probes of the late-stage dynamics of heavy-ion collisions, offering valuable insight into the properties of the hadronic phase~\cite{Brown:1991kk, Bleicher:2002dm, Torrieri:2001ue, Johnson:1999fv, Markert:2008jc, Ilner:2017tab, Shapoval:2017jej}.

% ADD SOME MORE TEXT HERE
Recent studies from the ALICE experiment have highlighted the importance of measuring the resonance production to study the interactions during the hadronic phase, not only in lead-lead (Pb--Pb) collisions but also in smaller collision systems~\cite{ALICE:2016sak, ALICE:2019etb, ALICE:2019xyr, ALICE:2023edr, ALICE:2023ifn}.
%Add something about alice result
For example, the studies of $\mathrm{K^{*}(892)^0}$ and $\phi(1020)$ mesons production measurement in proton--proton (pp) collisions at $\sqrt{s}=13$ TeV and proton--lead (p--Pb) collisions at $\sqrt{s_\mathrm{NN}}=5.02$ TeV reveal that the yield ratio of $\mathrm{K^*(892)^0}$ to charged kaons decreases with increasing multiplicity~\cite{ALICE:2019etb, ALICE:2016sak}, showing a similar trend as that in large collision systems~\cite{ALICE:2023edr}.
Also, a detailed investigation using transport model, EPOS3~\cite{Werner:2007bf, Werner:2013tya} with UrQMD~\cite{Bleicher:1999xi, Bass:1998ca}, has been performed in Ref.~\cite{knospe2021hadronic} to interpret the experimental data on resonance production.
These studies emphasize that hadronic interactions significantly modify the yields and momentum distributions of short-lived resonances, depending on their lifetimes and interaction cross sections. 
However, unlike p--Pb and Pb--Pb collisions, the interactions during the hadronic phase in pp collisions have not been deeply studied yet.

Recent observations in high-multiplicity pp collisions have revealed signatures reminiscent of QGP-like behavior~\cite{Nagle:2018nvi}. These include long-range azimuthal correlations, strangeness enhancement, and collective flow patterns, which were previously thought to be unique to large collision systems. Such findings have sparked significant interest in understanding whether a deconfined medium, similar to the QGP, can also emerge in small systems like pp collisions under extreme conditions. In addition to investigating the QGP formation at the partonic phase level, studying the hadronic phase will also be beneficial for a comprehensive understanding of the properties of the medium produced in pp collisions and its time evolution.

Therefore, this paper investigates how the hadronic interaction, especially the rescattering effect, impacts the resonance production in pp collisions at $\sqrt{s}=13$ TeV and p--Pb collisions at $\sqrt{s_\mathrm{NN}}=5.02$ TeV using the PYTHIA8 model, by switching on and off the hadronic rescattering effect implemented in the model~\cite{Sjostrand:2020gyg, Bierlich:2021poz, Bierlich:2022pfr}.
For the rescattering on option, both cases with and without inelastic scattering are studied.
As the lifetime of the hadronic phase in small systems may be shorter than 4 fm/$c$ or less~\cite{ParticleDataGroup:2024cfk, ALICE:2019xyr}, we will compare the particle yield ratios of $\rho(770)^0$, $\mathrm{K^*(892)^0}$, and $\mathrm{\phi(1020)}$ mesons which have the lifetime of 1.3 fm/$c$, 4.16 fm/$c$, and 46.3 fm/$c$, respectively.
This comparison will provide a rough constraint on the lifetime of the hadronic phase in each collision system.

\section{\label{sec:models}PYTHIA8 model}

PYTHIA8 is an event generator widely used for high-energy pp collisions, but also events of different collision systems can be generated by using the Angantyr model~\cite{Bierlich:2018xfw}.
This event generator incorporates both hard and soft interactions for jets and underlying events, and the default Monash tune provides a reasonable description of soft particle production in pp collisions~\cite{Skands:2014pea}.
However, as partonic and hadronic interaction is not supported in the Monash tune, this study uses PYTHIA 8.312 version which includes hadronic rescattering option~\cite{10.21468/SciPostPhysCodeb.8}.
In PYTHIA 8.312, hadronic rescattering is modeled by allowing the produced hadrons to scatter with nearby hadrons before they decouple, as the system expands and the density decreases.
%PYTHIA 8.312 describes the hadronic rescattering by assuming that the produced hadrons can undergo scattering with nearby hadrons each other until the hadrons get free as the fragmenting system expand enough. 
Each hadron pair is evaluated to determine whether it fulfills the interaction criteria. 
If so, the interaction time is recorded in a time-order list. 
Unstable particles may also decay during the rescattering phase, and the decay times are inserted into the list.
Unless a particle is already involved in another interaction or decay, its scattering and decay processes proceed according to the chronological order defined in the time-order list.
%Unless the particles involved in interaction or decay, the scattering or decay is proceeded in chronological order.
Resulting new hadrons are again evaluated for possible interactions or decays, and any valid events are added to the time-order list.
This process is repeated until no further interactions remain.

\section{Analysis procedure}
\label{sec:ana}
\subsection{Event and particle selections}

This analysis aims to investigate the production yield ratio of $\rho(770)^0$, $\mathrm{K^*(892)^0}$ and $\mathrm{\phi(1020)}$ mesons in pp collisions at $\sqrt{s}=13$ TeV and p--Pb collisions at $\sqrt{s_\mathrm{NN}}=5.02$ TeV. Simulations are performed using the PYTHIA8 event generator with and without the hadronic rescattering on option, as well as inelastic scattering on and off options when rescattering option is on.
For pp collisions, we generated non-diffractive events with an ``\texttt{SoftQCD:nonDiffractive}'' option and further required that events which have at least one charged particle in the pseudorapidity range $|\eta|<1.0$ region, which is denoted as INEL\ $>$\ 0 to make it comparable with ALICE results.
Also the same event selection criteria as the ALICE experiment were used for both pp and p--Pb collisions, which require at least one charged particle in both V0A and V0C~\cite{ALICE:2008ngc, ALICE:2013axi} acceptance. The pseudorapidity coverage of V0A and V0C is $2.8<\eta<5.1$ and $-3.7<\eta<-1.7$, respectively.
The multiplicity percentiles are estimated by V0M for pp collisions and V0A (Pb-going direction) for p--Pb collisions, where V0M is the sum of the charged particles in both the V0A and V0C acceptance.
The multiplicity percentile classes used are as follows: 0--5\%, 5--10\%, 10--25\%, 25--40\%, 40--60\%, 60--100\%.
For track selection, we considered all charged pions ($\pi^{\pm}$) and kaons ($\mathrm{K^{\pm}}$) for further analysis to reconstruct resonance particles. 
All final-state particles from PTYHIA8 were analyzed by applying analysis-level selections as well as resonance reconstruction, in a manner similar to the ALICE experimental procedure.
%We limited the decay time to 10 mm/$c$, excluding particles from weak decay.

\subsection{Yield extraction}

\begin{figure*}[tbh]
    \centering
    \includegraphics[width=0.32\textwidth]{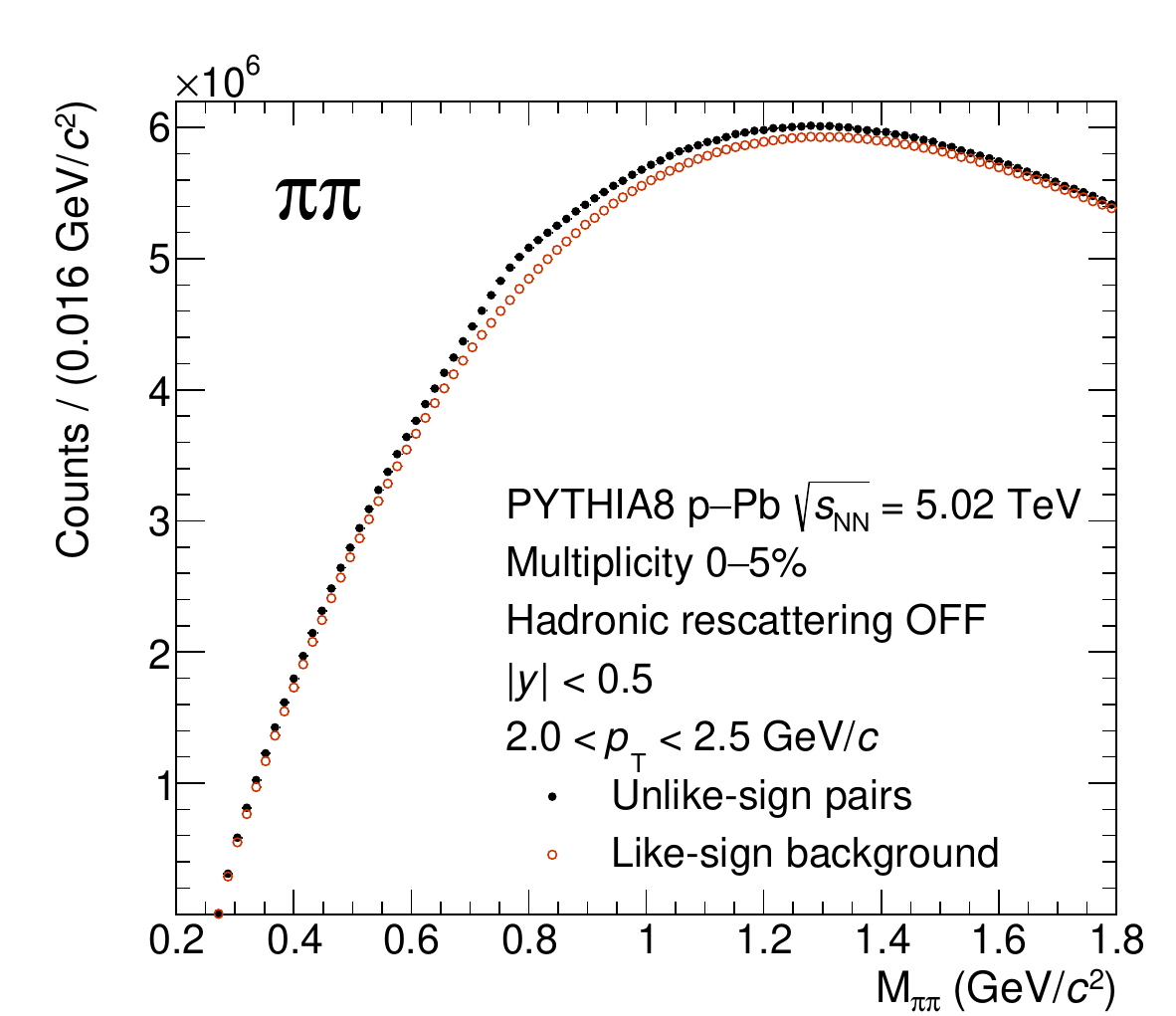}
    \includegraphics[width=0.32\textwidth]{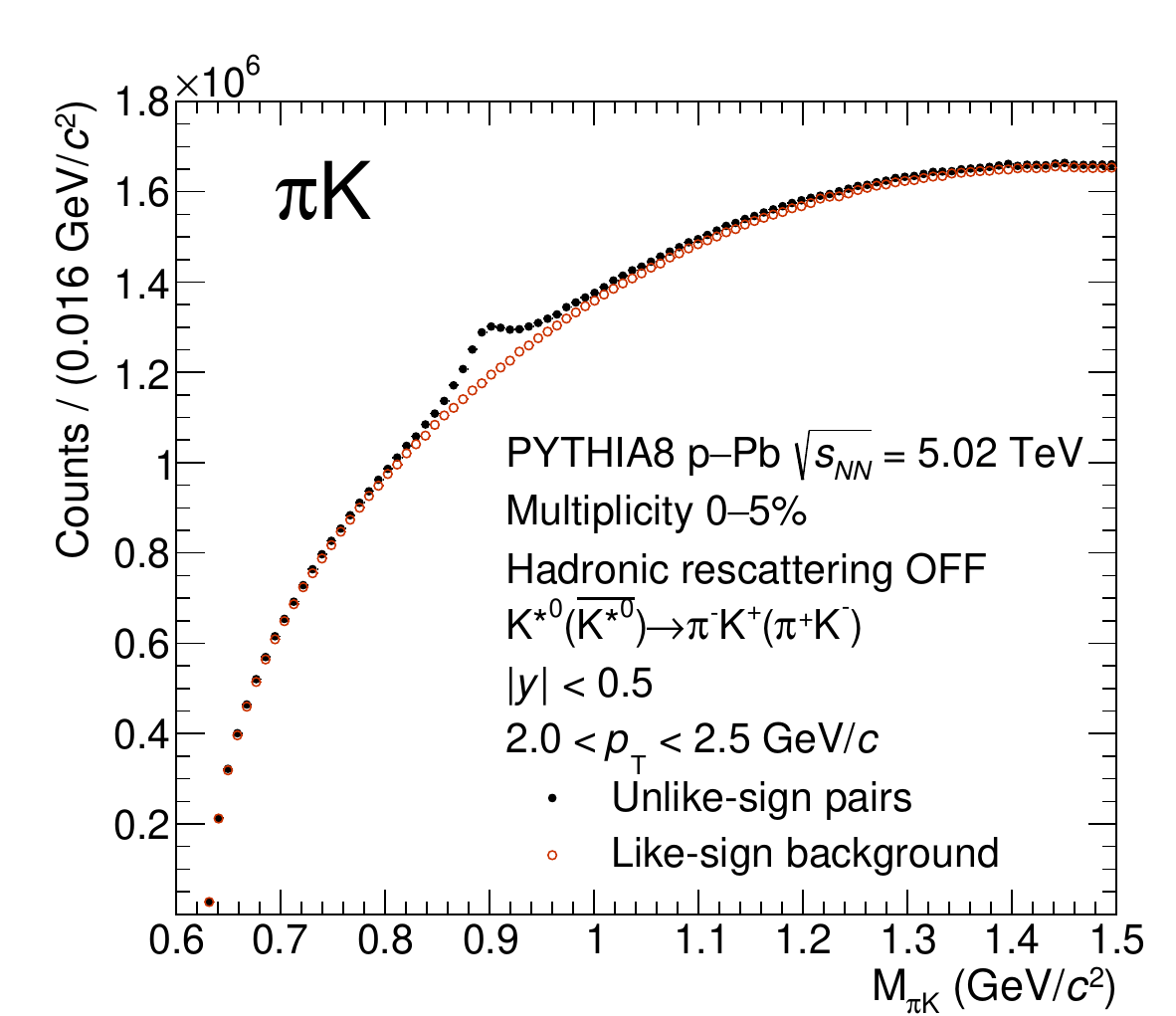}
    \includegraphics[width=0.32\textwidth]{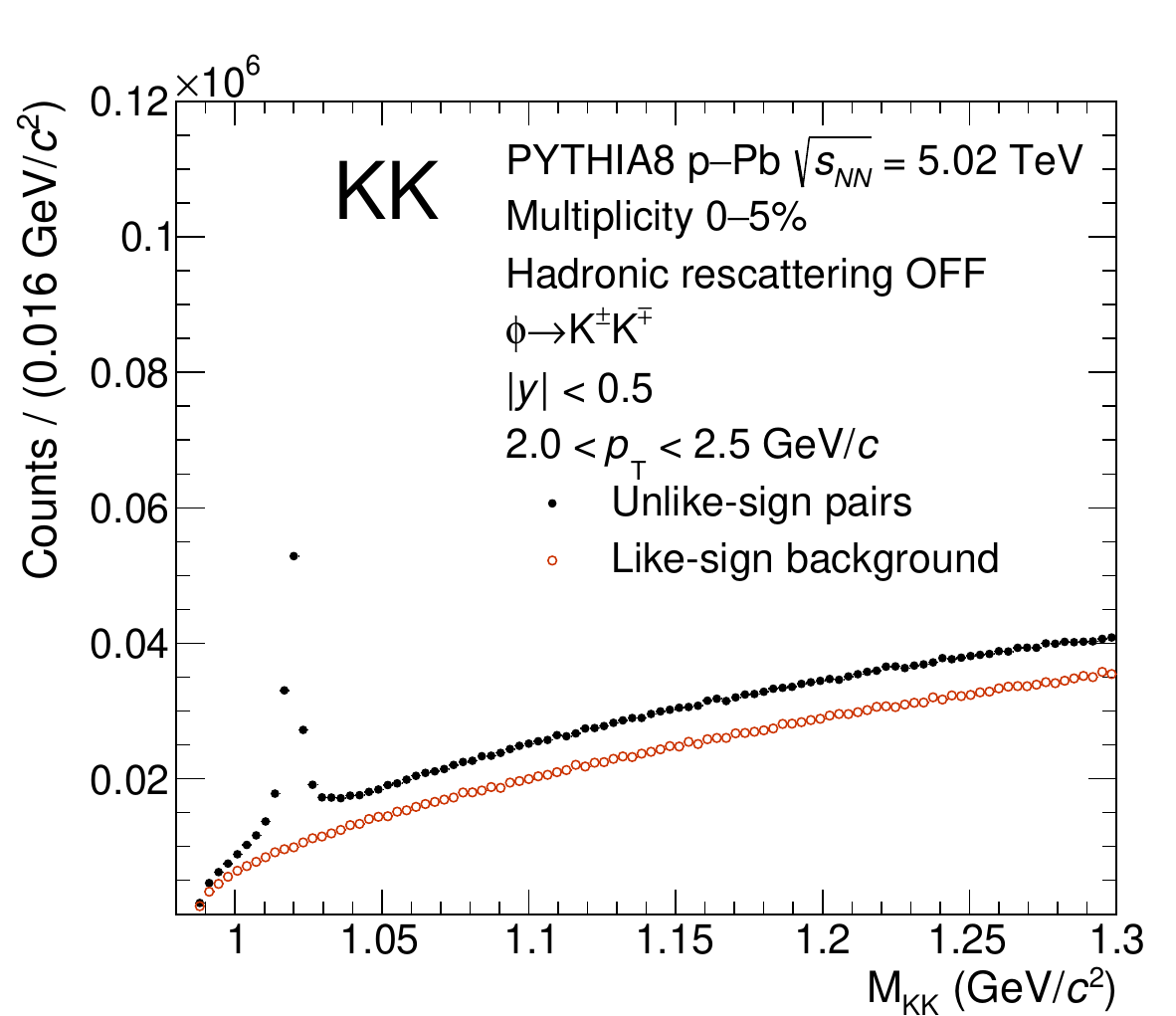}
    \includegraphics[width=0.32\textwidth]{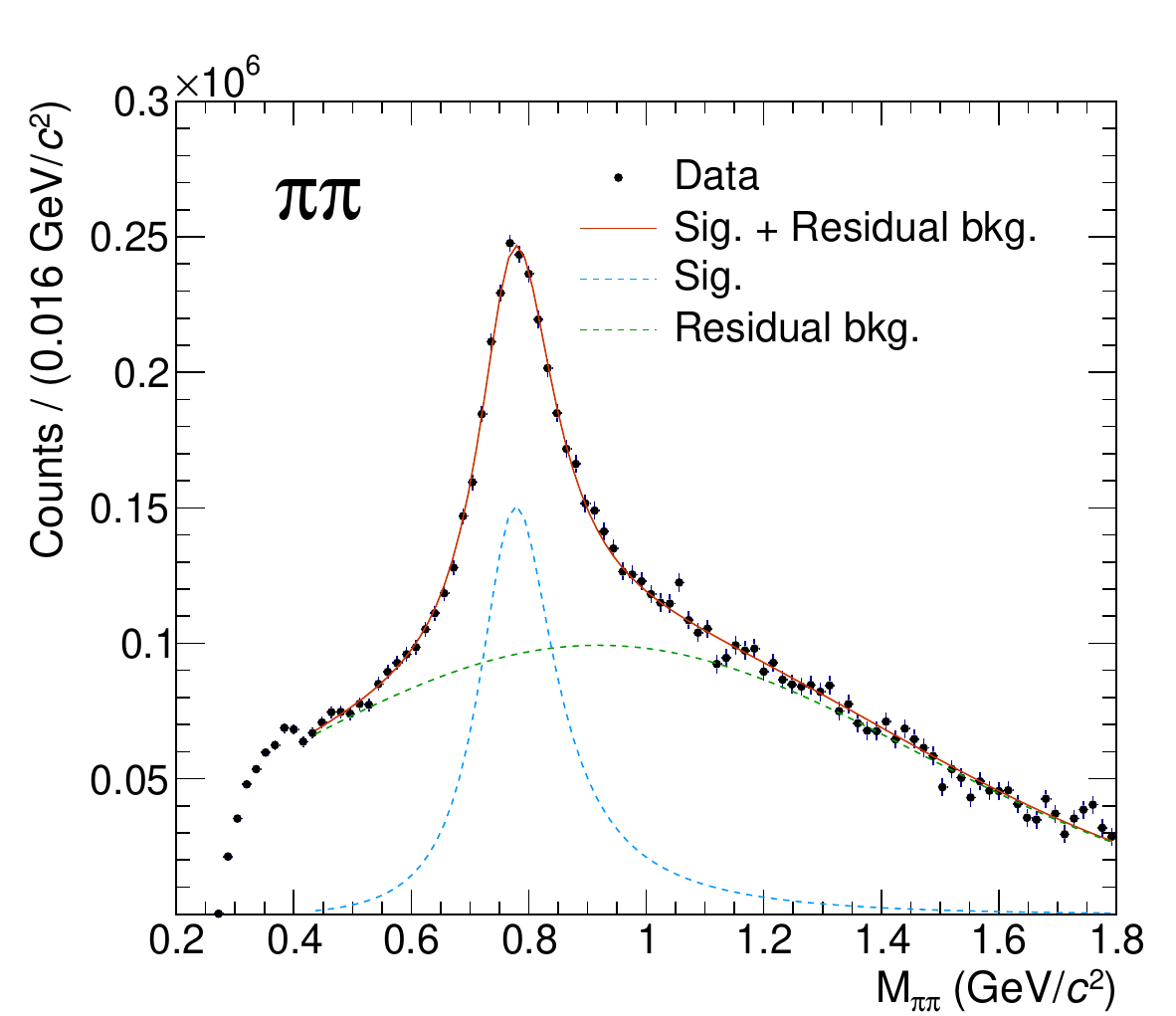}
    \includegraphics[width=0.32\textwidth]{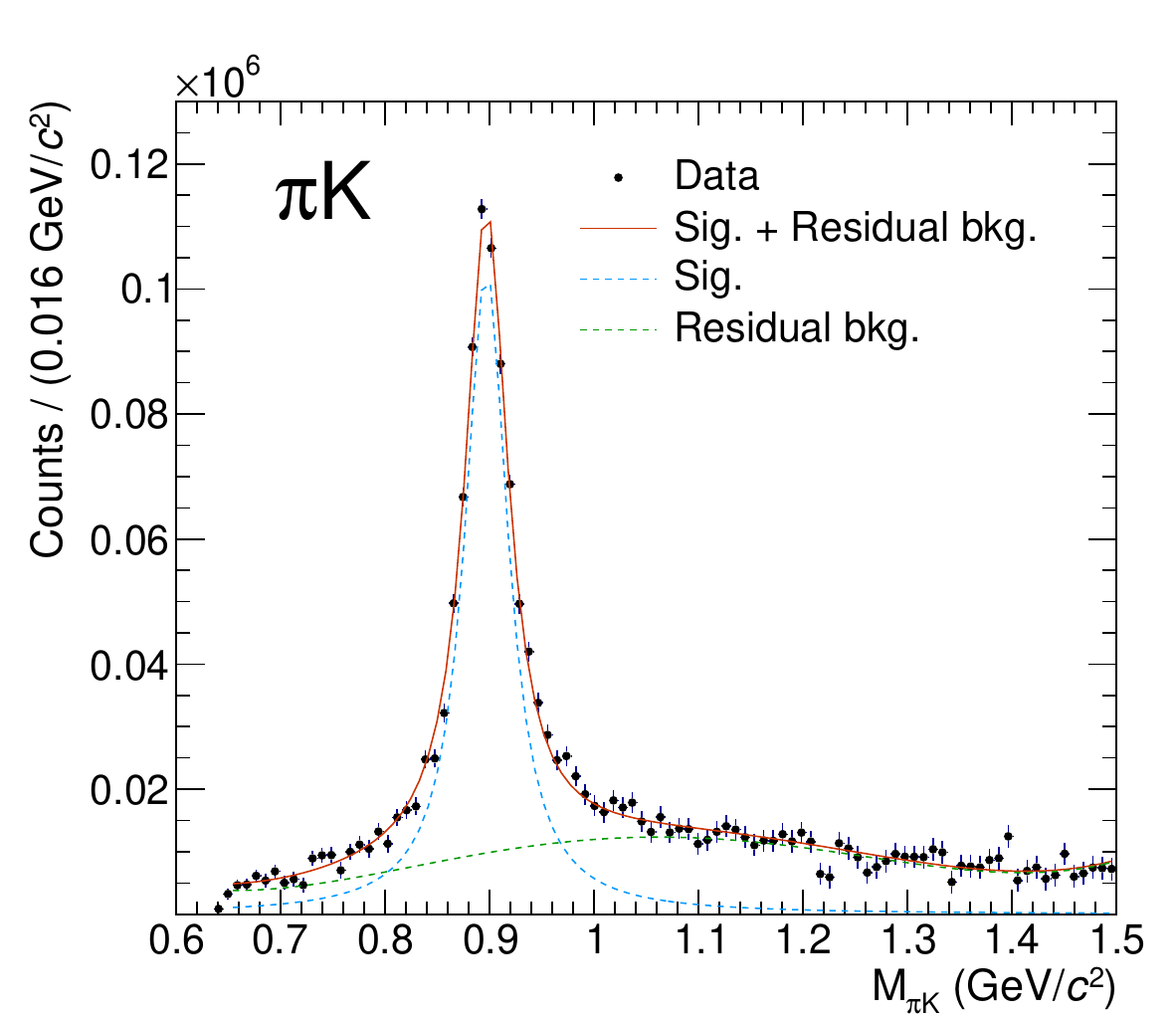}
    \includegraphics[width=0.32\textwidth]{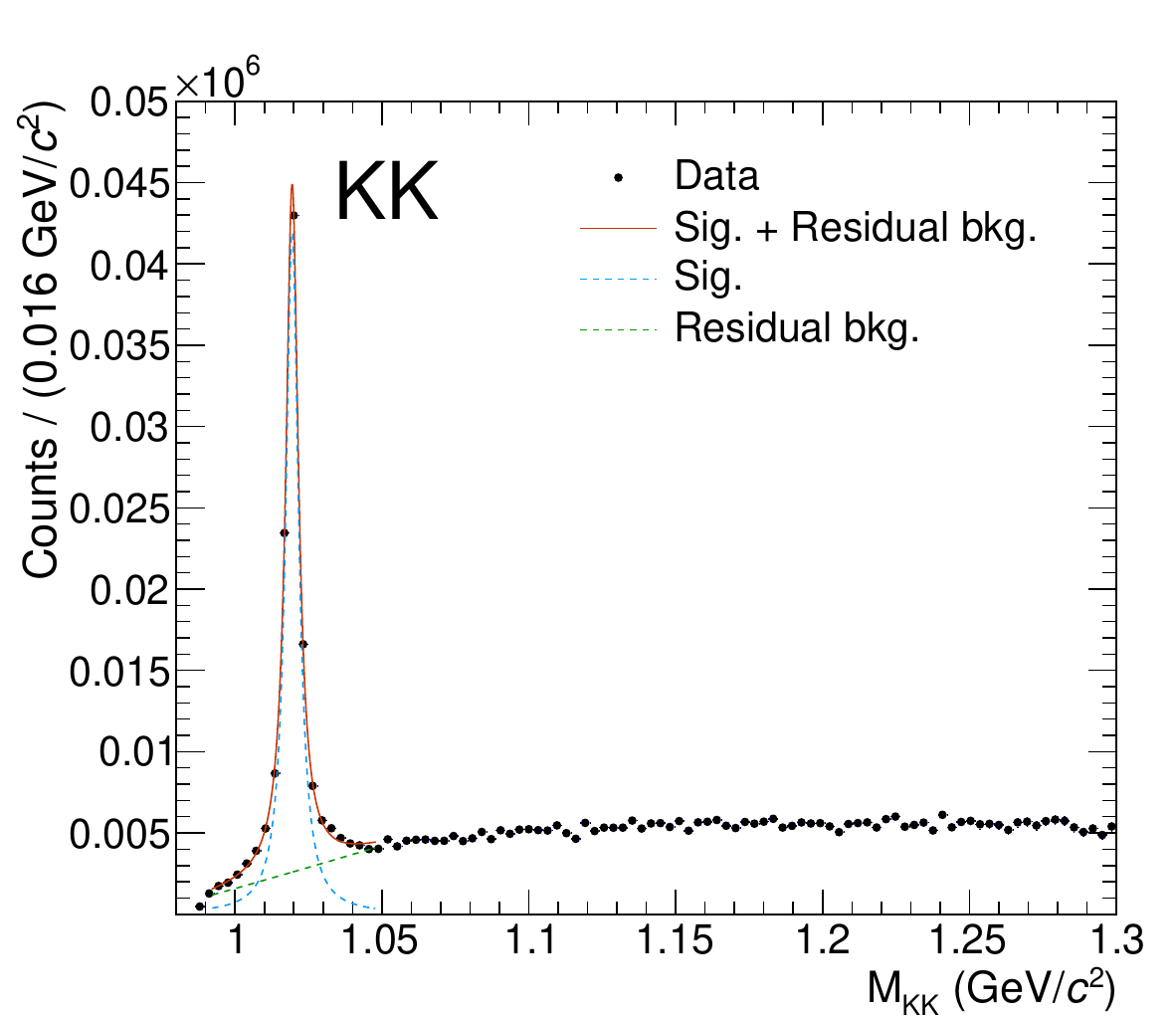}
    \caption{The invariant mass distribution for $\rho(770)^0$ (left), $\mathrm{K^{*}(892)^{0}}$ (middle), and $\phi(1020)$ mesons (right) in multiplicity 0--5\% in $2.0<p_\mathrm{T}<2.5$ GeV/$c$ before (top) and after (bottom) the background subtraction, when the rescattering option is off. In the bottom panels, fit results with signal and background functions are also presented.}
    \label{fig:Fit_all}
\end{figure*}

The yield of $\rho(770)^0$, $\mathrm{K^{*}(892)^{0}}$, and $\phi(1020)$ resonances are reconstructed through the invariant mass analysis.
The hadronic decay channels of these particles are as follows: 
\begin{itemize}
    \item $\rho(770)^0\rightarrow\pi^+\pi^-$,
    \item $\mathrm{K^{*}(892)^{0}}(\overline{\mathrm{K}}^*(892)^0)\rightarrow\mathrm{\pi^- K^{+}(\pi^+ K^{-})}$,
    \item $\phi(1020)\rightarrow\mathrm{K^+ K^-}$.
\end{itemize}
%In the case of $\rho(770)^0$ meson analysis, $\mathrm{K_s^0}$ and $\omega$ mesons also decay into $\pi^{+}\pi^{-}$ pairs, making it difficult to extract raw yields of $\rho(770)^0$. Therefore, $\pi^{\pm}$ from $\mathrm{K_s^0}$ and $\omega$ decays are removed by checking the decay history in PYTHIA8.
In the analysis of the $\rho(770)^0$ meson, $\mathrm{K_S^0}$ and $\omega$ mesons also decay into $\pi^{+}\pi^{-}$ pairs, complicating the extraction of raw $\rho(770)^0$ yields. To address this, $\pi^{\pm}$ originating from $\mathrm{K_S^0}$ and $\omega$ decays are excluded by tracing the decay history in PYTHIA8.
For $\mathrm{K^{*}(892)^{0}}$ and $\phi(1020)$ mesons, there is no such requirement in the analysis procedure.
For the combinatorial background, it is estimated using like-sign pairs ($\pi^{\pm}\pi^{\pm}$ for $\rho(770)^0$, and $\pi^{\pm}\mathrm{K}^{\pm}$ for $\mathrm{K^{*}(892)^{0}}$, and $\mathrm{K^{\pm}K^{\pm}}$ for $\phi(1020)$) by forming pairs from particles generated in the same events.
The background has been subtracted from the invariant mass distribution of the unlike-sign pairs, yet the residual background still remains.
Therefore, the fit procedures are conducted on the background-subtracted invariant mass distribution using a combined function that describes both the residual background and the signal peak to extract the signal of the resonances.
The fit regions for each particle are $0.45<m_{\pi\pi}<1.80$ GeV/$c^2$ for $\rho(770)^0$, $0.65<m_{\pi \mathrm{K}}<1.50$ GeV/$c^2$ for $\mathrm{K^{*}(892)^{0}}$, and $0.99<m_{\mathrm{KK}}<1.055$ GeV/$c^2$ for $\phi(1020)$ meson.

Fig.~\ref{fig:Fit_all} shows examples of the invariant mass distribution before (top) and after (bottom) the combinatorial background subtraction for $\rho(770)^0$, $\mathrm{K^{*}(892)^{0}}$, and $\phi(1020)$ mesons in transverse momentum (\pt) region $2.0<p_{\rm T}<2.5$ GeV/$c$ and $|y|<0.5$. Events are in the 0--5\% V0A multiplicity class from PYTHIA8 p--Pb events at $\sqrt{s_\mathrm{NN}}=5.02$ TeV with the hadronic rescattering off option.
Both unlike-sign and like-sign invariant mass distributions are presented in the top panels, and the fit results of the background-subtracted distributions, using functions for signal and residual background described below, are shown.
For the fitting procedure of the $\rho(770)^0$ meson, we followed the method described in Ref.~\cite{ALICE:2018qdv}. The mass and width of the signal peak are extracted from the relativistic Breit–Wigner function with phase space correction:
\begin{align}
\frac{dN_{\mathrm{sig}}}{dM_{\pi\pi}} = &\frac{AM_{\pi\pi}M_0\Gamma(M_{\pi\pi})}{(M_0^2-M^2_{\pi\pi})^2+M_0^2\Gamma^2(M_{\pi\pi})}\nonumber\\
& \times \left[\frac{M_{\pi\pi}}{\sqrt{M_{\pi\pi}^2+p_\mathrm{T}^2}}\exp\left( -\frac{\sqrt{M_{\pi\pi}^2+p_\mathrm{T}^2}}{T} \right)\right],
\end{align}
\label{eq:rBWwPS}
where $A$ is a normalization constant, and $M_0$ and $\Gamma(M_{\pi\pi})$ are the mass and width of the $\rho(770)^0$ resonance, which $\Gamma(M_{\pi\pi})$ depends on the mass from the invariant mass distribution $m_{\pi\pi}$ and the width of resonance $\Gamma_0$, and the rest mass of charged pion ($m_\pi$):
\begin{align}
\Gamma(M_{\pi\pi})=\left( \frac{M_{\pi\pi}^2-4m_\pi^2}{M_0^2-4m_\pi^2}\right)^{3/2}\times\Gamma_0\times M_0/M_{\pi\pi}.
\end{align}
\label{eq:rBWwPSGamma}
The temperature $T$ is fixed to 160 MeV, which is approximately the same as the chemical freeze-out temperature~\cite{ALICE:2018qdv}.
For the residual background description of $\rho(770)^0$ meson, unlike-sign pairs are selected which originated from different mother particles, with a fit function:
\begin{align}
\frac{dN_{bkg}}{dM_{\pi\pi}}=B\times\left(M_{\pi\pi}-2m_\pi\right)^n\times\exp(a+bM_{\pi\pi}+cM_{\pi\pi}^2).
\end{align}
\label{eq:rBWwPSbkg}

For $\mathrm{K^{*0}}$ and $\phi$ meson fitting, the Breit-Wigner function is used to describe the signal peaks~\cite{ALICE:2017ban}:
\begin{align}
\frac{dN_{\mathrm{sig}}}{dM_{\pi\mathrm{K}(\mathrm{KK})}} = \frac{A}{(M_{\pi\mathrm{K}(\mathrm{KK})}-M_0)^2+\Gamma_0^2/4}.
\end{align}
\label{eq:rBWwPS}
Each component corresponds with the notation described above.
For the residual background of $\mathrm{K^{*}(892)^{0}}$ and $\phi(1020)$, fourth-order polynomial and second-order polynomial functions are used, respectively, and the parameters are determined in the fit procedure. 
After the fit procedure is applied, the raw yield of resonances in each multiplicity and $p_{\rm T}$ bin is calculated by integrating the signal function.

To study the effect of the hadronic rescattering and inelastic collisions on the invariant mass distributions of each particle, the position of mean and the width of the signal functions in the highest multiplicity, 0--5\% and the lowest multiplicity, 60--100\% are compared in p--Pb collisions for each configuration.
\begin{figure*}[tbh]
    \centering
    \includegraphics[width=0.99\textwidth]{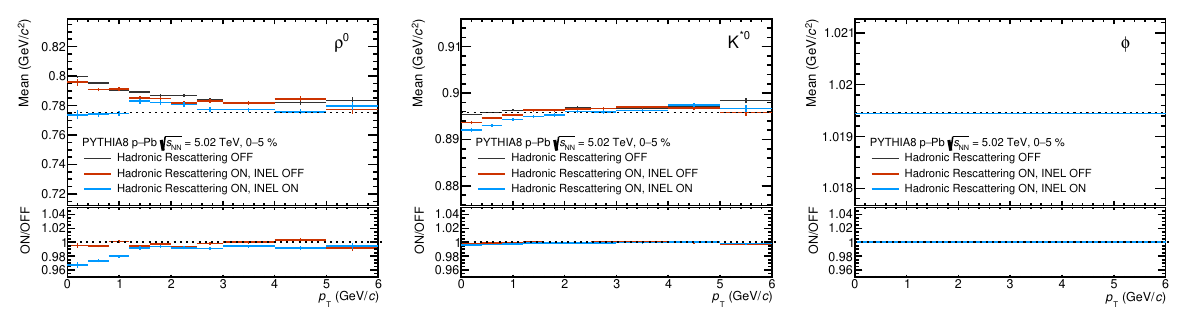}
    \includegraphics[width=0.99\textwidth]{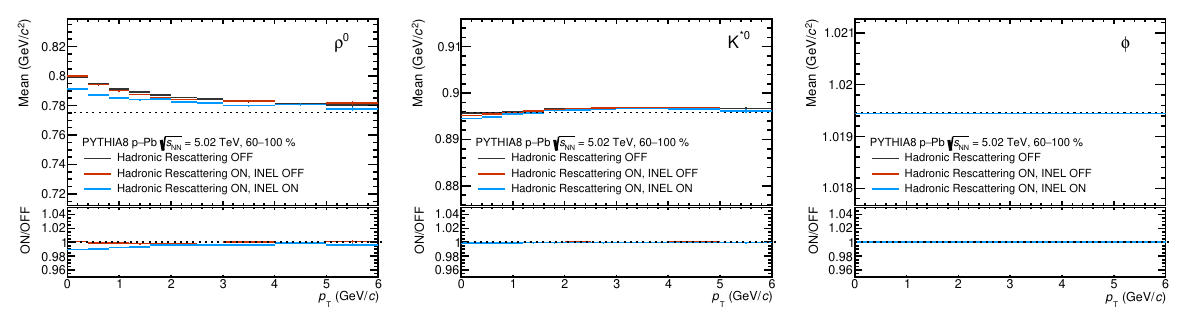}
    \caption{Mean values extracted from fit functions as a function of $p_\mathrm{T}$ in p--Pb collisions at $\sqrt{s_\mathrm{NN}}=5.02$ TeV for the 0--5\% (top) and 60--100\% (bottom) multiplicity classes. Results are shown for $\rho(770)^0$ (left), $\mathrm{K^{*}(892)^{0}}$ (middle), and $\phi(1020)$ (right) mesons. The PDG reference values are indicated by dashed lines.}
    \label{fig:Mean_pPb}
\end{figure*}
%In Fig.~\ref{fig:Mean_pPb}, the mean position of the signal function of each particle is presented. For both 0--5\% and 60--100\%, the configuration dependence of the mean position is not seen for $\mathrm{K^{*}(892)^{0}}$ and $\phi(1020)$ mesons, also showing similar values with PDG values. For $\rho(770)^0$, hadronic rescattering off and rescattering on with inelastic scattering off options do not show significant difference, however show mean position shift to the higher mass with decreasing $p_\mathrm{T}$, while hadronic rescattering on with inelastic scattering on option shows consistent result with the one from PDG across all the $p_\mathrm{T}$ bins for 0--5\% multiplicity. In the case of 60--100\% events, hadronic rescattering on with inelastic scattering on shows closer mean position with both hadronic rescattering off and inelastic scattering on compared to 0--5\%, however shows discrepancy between PDG value for all configurations.
In Fig.~\ref{fig:Mean_pPb}, the mean position of the signal function for each particle is shown. For both the 0--5\% and 60--100\% multiplicity classes, no configuration dependence is observed for the $\mathrm{K^{*}(892)^{0}}$ and $\phi(1020)$ mesons, and the mean positions are consistent with the PDG values. For the $\rho(770)^0$ meson, the results with hadronic rescattering off and with rescattering on (inelastic scattering off) do not differ significantly; however, both configurations exhibit a mean position shifted to higher mass with decreasing $p_\mathrm{T}$. In contrast, the hadronic rescattering on with inelastic scattering on configuration yields mean positions consistent with the PDG values across all $p_\mathrm{T}$ bins in the 0--5\% multiplicity class. In the 60--100\% events, the hadronic rescattering on with inelastic scattering on configuration shows mean positions closer to those from the rescattering-off and inelastic-scattering-off cases than in 0--5\%. Nevertheless, discrepancies with the PDG values remain for all configurations.

\begin{figure*}[tbh]
    \centering
    \includegraphics[width=0.99\textwidth]{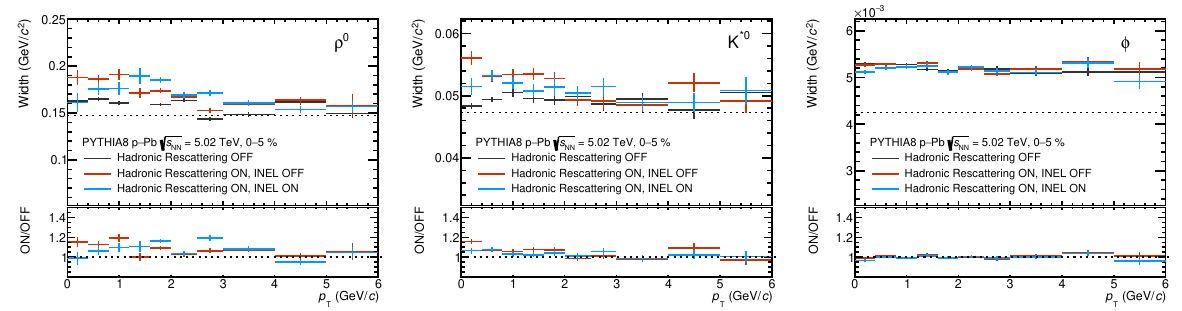}
    \includegraphics[width=0.99\textwidth]{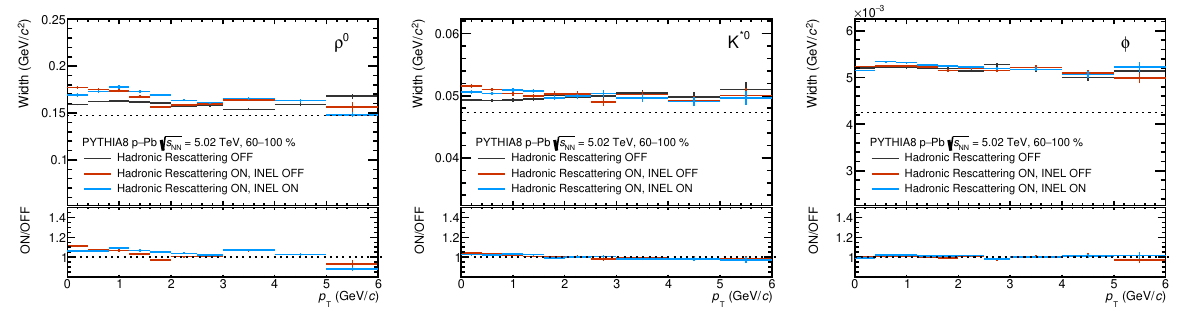}
    \caption{Width values extracted from fit functions as a function of $p_\mathrm{T}$ in p--Pb collisions at $\sqrt{s_\mathrm{NN}}=5.02$ TeV for the 0--5\% (top) and 60--100\% (bottom) multiplicity classes. Results are shown for $\rho(770)^0$ (left), $\mathrm{K^{*}(892)^{0}}$ (middle), and $\phi(1020)$ (right) mesons. The PDG reference values are indicated by dashed lines.}
    \label{fig:Width_pPb}
\end{figure*}
In addition, the widths of the signal shapes obtained from invariant mass fits were compared for the three PYTHIA configurations. Figure~\ref{fig:Width_pPb} shows the $p_\mathrm{T}$ dependence of the width values in 5.02 TeV p--Pb collisions at 0--5\% and 60--100\% centralities. The lower panels of each figure display the ratio between the results with the hadronic rescattering on and off options. For the $\phi(1020)$ meson (right panels), the results with rescattering on and off showed no significant differences, consistent with the behavior observed in the mean values. In the case of the $\mathrm{K^{*}(892)^{0}}$, however, the trend differs from that of the mean values. In 0--5\% p--Pb collisions, the width with hadronic rescattering on is about 5--10\% larger than that with rescattering off in the low-$p_\mathrm{T}$ region ($p_\mathrm{T} < 2$ GeV/$c$). For 60--100\% p--Pb collisions, the difference is much smaller, less than 5\% at $p_\mathrm{T} < 1$ GeV/$c$.
For the $\rho(770)^0$ meson, the results are similar to those of the $\mathrm{K^{*}(892)^{0}}$, with the width being larger for rescattering on than for rescattering off. Moreover, the difference between the two configurations is more pronounced for the $\rho(770)^0$ than for the $\mathrm{K^{*}(892)^{0}}$. Due to the uncertainties in the width values, however, it was not possible to clearly distinguish the effect of turning the inelastic scattering option on or off.

\begin{figure*}[tbh]
    \centering
    \includegraphics[width=0.99\textwidth]{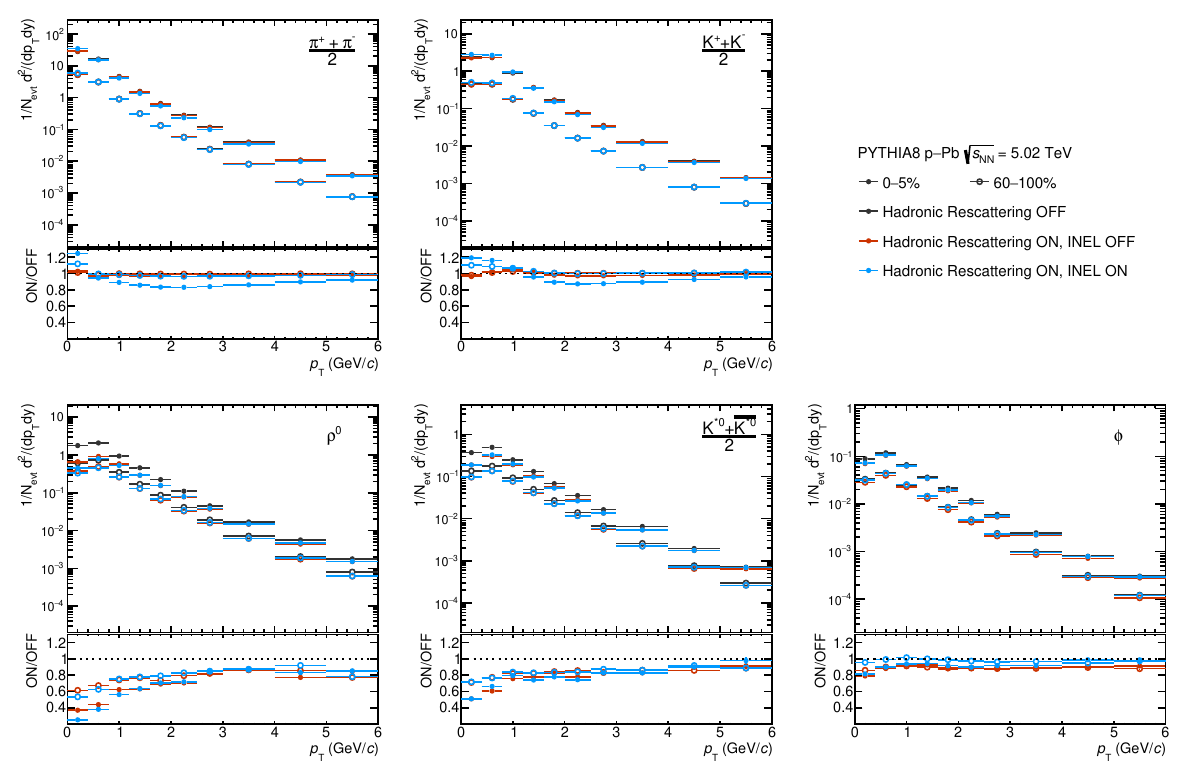}
    \caption{Comparison of $p_\mathrm{T}$ spectra for the three configurations—hadronic rescattering on with inelastic scattering on, hadronic rescattering on with inelastic scattering off, and hadronic rescattering off—for stable particles (top) and resonances (bottom) in the 0--5\% and 60--100\% multiplicity classes. The lower panels show the ratios of spectra with hadronic rescattering on to those with rescattering off.}
    \label{fig:pTspectra}
\end{figure*}

%The rescattering effect not only affect on the mean and width of the particles, but also on the $p_\mathrm{T}$ spectra of the particles. Fig.~\ref{fig:pTspectra} shows the $p_\mathrm{T}$ spectra of each resonance and stable particle in 0--5\% and 60--100\% events for all three configurations.
%If we focus on the stable particles, $\pi^\pm$ and $\mathrm{K^\pm}$ first, the hadronic rescattering does not show significant effect in 60--100\%, showing similar value with hadronic rescattering off option apart from the very low $p_\mathrm{T}$ region. The low $p_\mathrm{T}$ region shows higher value for hadronic rescattering on with inelastic scattering on option compared to the others. 
%For 0--5\%, hadronic rescattering on with inelastic on option shows enhancement in low $p_\mathrm{T}$ region, however shows suppression in intermediate $p_\mathrm{T}$ compared to the other configurations.
%Unlike the stable particles, the $\rho(770)^0$ and $\mathrm{K^*(892)^0}$ show strong suppression especially in low $p_\mathrm{T}$ region for both 60--100\% and 0--5\%, where the degree of suppression is stronger in 0--5\%. For $\phi(1020)$ meson, slight suppression is seen across all $p_\mathrm{T}$ region without showing any $p_\mathrm{T}$ dependence.

The rescattering effect influences not only the mean values and widths of resonance signal shapes but also the transverse momentum ($p_\mathrm{T}$) spectra of particles. Figure~\ref{fig:pTspectra} presents the $p_\mathrm{T}$ spectra of each resonance and stable particle in 0--5\% and 60--100\% events for all three PYTHIA configurations.
For stable particles, namely $\pi^\pm$ and $\mathrm{K^\pm}$, the effect of hadronic rescattering is minimal in 60--100\% events, where the spectra are nearly identical to those obtained with the rescattering-off configuration, except in the very low-$p_\mathrm{T}$ region. In this region, the hadronic rescattering on with inelastic scattering on configuration yields slightly higher values than the others. In 0--5\% events, the same configuration produces an enhancement at low $p_\mathrm{T}$, but a suppression at intermediate $p_\mathrm{T}$ compared to the other two configurations.
In contrast, the $\rho(770)^0$ and $\mathrm{K^{*}(892)^0}$ resonances exhibit strong suppression at low $p_\mathrm{T}$ in both central (0--5\%) and peripheral (60--100\%) events, with the suppression being more pronounced in the central case. For the $\phi(1020)$ meson, a mild suppression is observed across the entire $p_\mathrm{T}$ range, without any clear $p_\mathrm{T}$ dependence.

Since stable particles are not affected on the rescattering effect dominantly, we take the yield ratio between resonances and stable particles to study the interactions during the hadronic phase. This will be discussed more in the Results section.
Please note that from now on, the $(\pi^++\pi^-)/2$, $(\mathrm{K}^++\mathrm{K}^-)/2$ and $(\mathrm{K^{*0}+\overline{K^{*0}}})/2$ are noted as $\pi$, $\mathrm{K}$ and $\mathrm{\mathrm{K^{*0}}}$.

\section {Results}
\label{sec:results}
This section presents the results in three steps: first, the particle ratios are evaluated; second, the effects of hadronic rescattering are examined by comparing simulations with rescattering turned on and off, as well as inelastic scattering turned on and off when rescattering is on; finally, the lower limit of the lifetime of the hadronic phase is extracted based on the observed modifications.

\subsection{Particle ratios}

\begin{figure*}[htb]
    \centering
	\includegraphics[width=0.975\textwidth]{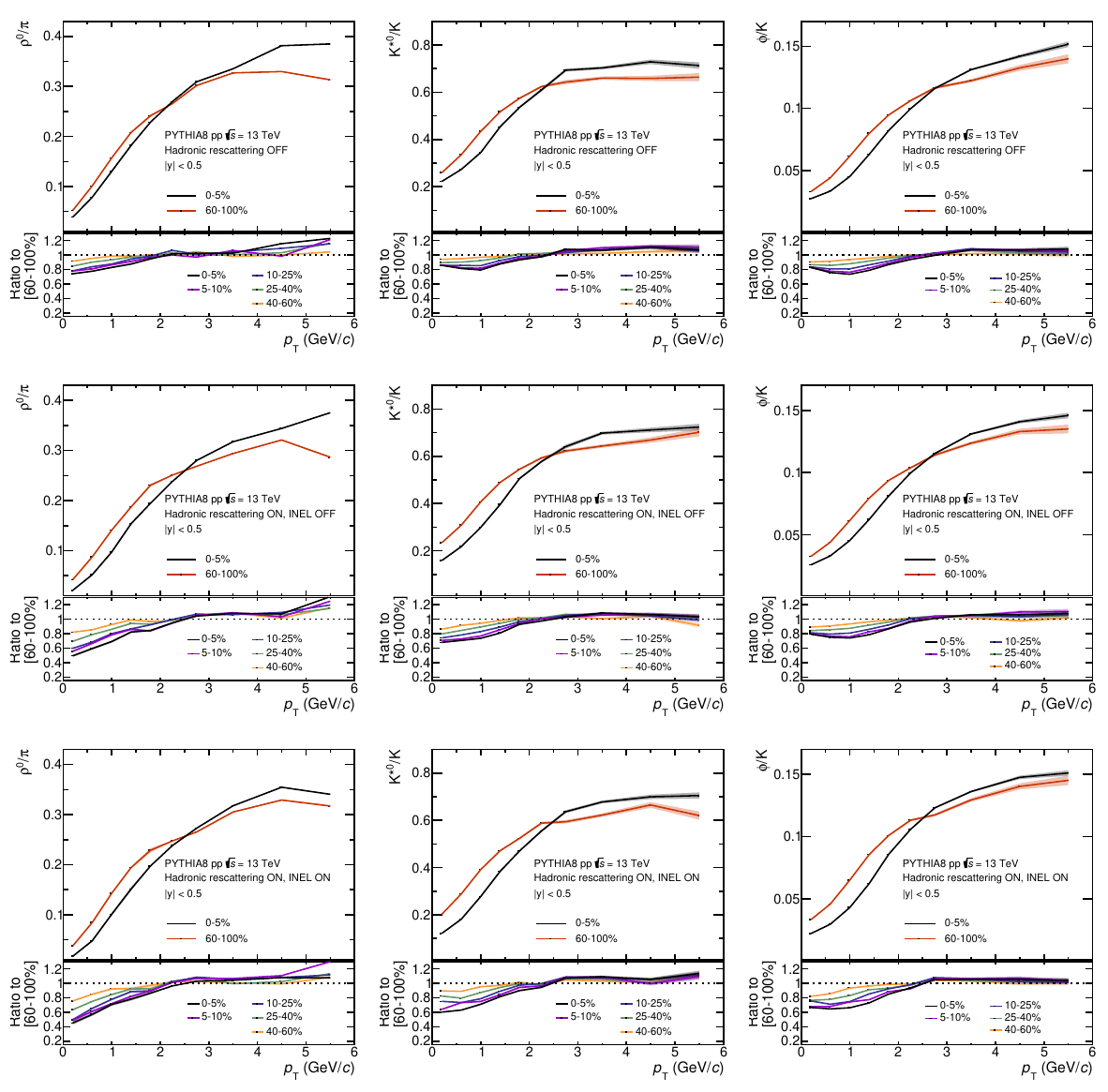} 
    \caption{The particle yield ratios between resonances and stable particles as a function of \pt\ in 0--5\% and 60--100\% multiplicity classes in pp collisions $\sqrt{s}=13$, which top figure represents hadronic rescattering off, middle figure for hadronic rescattering on with inelastic collision off, and the bottom figure for hadronic rescattering on with inelastic collision on. In each figure, the ratios of different multiplicity classes to the 60--100\% multiplicity class are shown.}
    \label{fig:YR_pp}
\end{figure*}

\begin{figure*}[htb]
    \centering
	\includegraphics[width=0.975\textwidth]{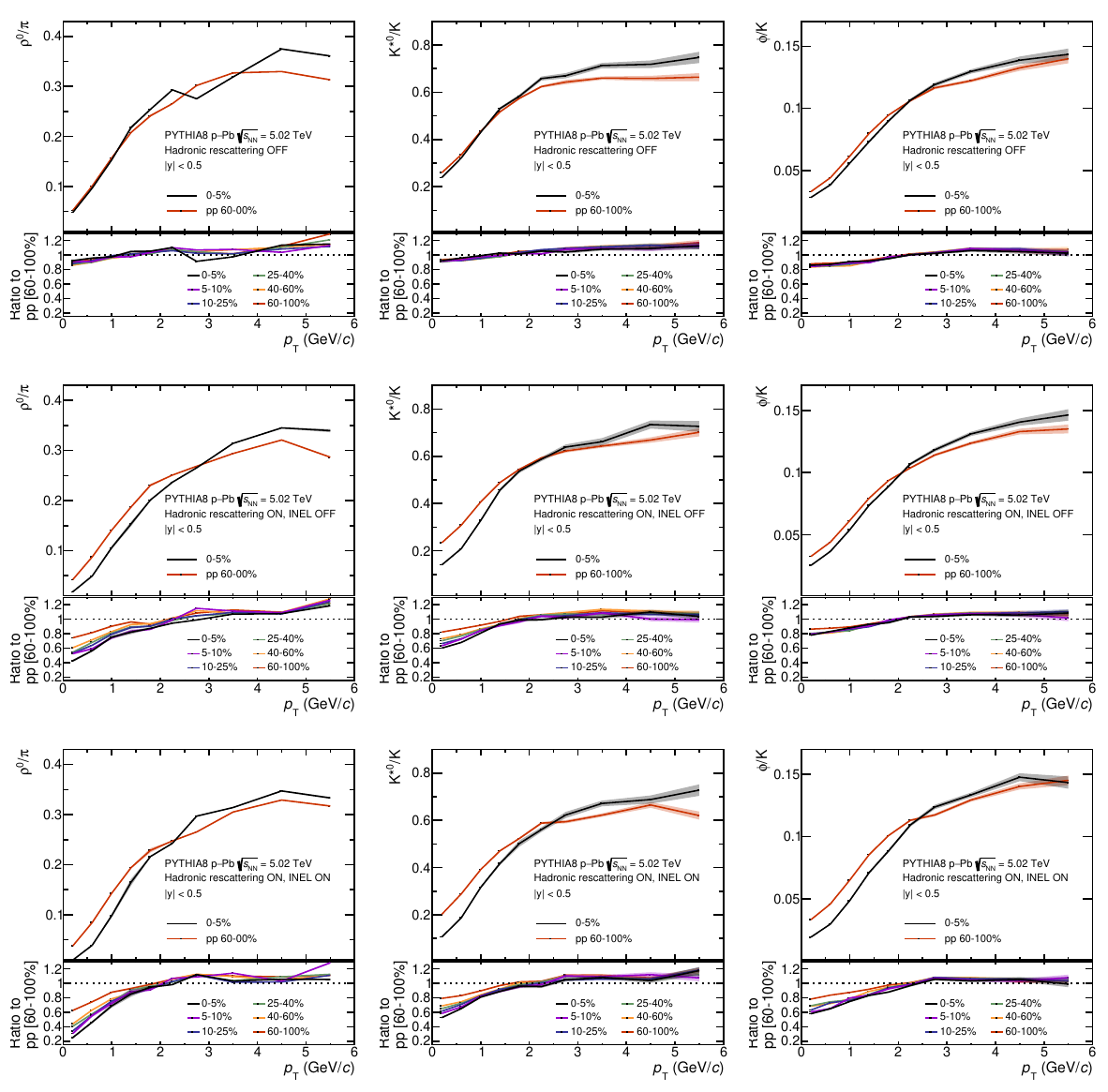} 
    \caption{The particle yield ratios between resonances and stable particles as a function of \pt\ in 0--5\% multiplicity class of p--Pb collisions at $\sqrt{s_{\rm NN}}=5.02$ and 60--100\% multiplicity class in pp collisions at $\sqrt{s}=13$, which top figure represents hadronic rescattering off, middle figure for hadronic rescattering on with inelastic collision off, and the bottom figure for hadronic rescattering on with inelastic collision on. In each figure, the ratios of different multiplicity classes of p--Pb collisions to the 60--100\% multiplicity class of pp collisions are shown.}
    \label{fig:YR_pPb}
\end{figure*}

The ratios between the resonances and stable particles with similar quark contents, $\rho^0/\pi$, $\mathrm{K^{*0}/K}$ and $\phi/\mathrm{K}$ are calculated as a function of \pt\ and shown in Fig.~\ref{fig:YR_pp} for pp collisions at \s\ = 13 TeV and in Fig.~\ref{fig:YR_pPb} for p--Pb collisions at \snn\ = 5.02 TeV generated using the PYTHIA8 event generator.
%$\rho(770)^0/(\pi^++\pi^-)$, $\mathrm{K^{*}(892)^{0}/(K^++K^-)}$, and $\phi(1020)/\mathrm{(K^++K^-)}$ are calculated as a function of \pt\ and shown in Fig.~\ref{fig:YR_pp} for pp collisions at $\sqrt{s}=13$ TeV and in Fig.~\ref{fig:YR_pPb} for p--Pb collisions at $\sqrt{s_\mathrm{NN}}=5.02$ TeV generated using the PYTHIA8 event generator. 
The top figures show the results without the hadronic rescattering option, and the middle figures are with hadronic rescattering but without inelastic scattering, and the bottom figures are with hadronic rescattering and inelastic scattering option.
In each figure, the ratios of different multiplicity classes of each collision to the 60--100\% multiplicity class from pp collisions are presented.

%For pp collisions, both rescattering on and off options, the ratio in the 60--100\% multiplicity class shows higher values compared to the ratio in the 0--5\% multiplicity class at \pt\ $<2.5$ GeV/$c$, and then 0–5\% takes over as \pt\ becomes higher. The double ratios, ratio to 60–100\% events, multiplicity hierarchy is observed in \pt\ $<2$ GeV/$c$ for both rescattering on and off, having the lowest ratio for the highest multiplicity class.
%In the case of rescattering off option, the double ratios are below unity in \pt\ $<2.5$ GeV/$c$, yet no significant difference is seen for three resonances apart from the lowest \pt\ interval, while rescattering on option shows that the double ratio of $\rho^0/(\pi^++\pi^-)$ reaches down to 0.4, while $\mathrm{K^{*0}/(K^++K^-)}$ and $\phi/\mathrm{(K^++K^-)}$ have the similar value of 0.6 and 0.7 for the lowest value, respectively.

For pp collisions, regardless of the hadronic rescattering setting (on or off), the ratio in the 60--100\% multiplicity class is higher than that in the 0--5\% class for \pt\ below 2 to 2.5 GeV/\textit{c}. However, as \pt\ increases beyond this point, the ratio of 0–5\% multiplicity class begins to dominate slightly. The double ratios—defined as the ratio normalized to that of the 60--100\% multiplicity class—exhibit a clear multiplicity hierarchy in the \pt\ region below 2 GeV/$c$ for both with and without using the hadronic rescattering option, and the lowest values are observed in the highest multiplicity class.
At \pt\ above 2 GeV/$c$, the ratio is slightly above the unity.

When the hadronic rescattering option is turned off, the double ratios remain below unity for $\pt < 2.5$ GeV/$c$, but only little difference is seen among the three resonances, except in the lowest \pt\ interval. In contrast, when the hadronic rescattering option is turned on, especially when the inelastic scattering is on, the double ratio for $\rho^0/\pi$ decreases significantly, reaching down to 0.4 at low-\pt. Meanwhile, $\mathrm{K^{*0}/K}$ and $\phi/\mathrm{K}$ exhibit slightly higher minimum values, around 0.6 to 0.7, respectively. This indicates a more substantial suppression effect for the $\rho(770)^0$ in the presence of hadronic rescatterings, consistent with its shorter lifetime compared to the other resonances.
%$\rho(770)^0/(\pi^++\pi^-)$ decreases significantly, reaching down to 0.4 at low-\pt. Meanwhile, $\mathrm{K^{*}(892)^{0}/(K^++K^-)}$ and $\phi(1020)/\mathrm{(K^++K^-)}$ exhibit slightly higher minimum values, around 0.6 and 0.7, respectively. This indicates a more substantial suppression effect for the $\rho(770)^0$ in the presence of hadronic rescatterings, consistent with its shorter lifetime compared to the other resonances.

In p--Pb collisions shown in Fig.~\ref{fig:YR_pPb}, the same feature as pp collisions is seen for the particle ratios with the hadronic rescattering on options, however for without the hadronic rescattering option, the yield ratios from two multiplicities (note that the low multiplicity class is the 60--100\% from pp collisions) show comparable value in $\pt<2$ GeV/$c$.
Therefore, unlike the double ratios with the hadronic rescattering options, the multiplicity hierarchy is not observed for the hadronic rescattering off option, where the double ratios are calculated by taking the particle yield ratio of 60--100\% events in pp collisions as the denominator.
Similar to pp collisions, the double ratios of all three resonances show similar values for the hadronic rescattering off option. With the hadronic rescattering on options, a difference is observed between the double ratios of each particle, with a lower value compared to that of pp collisions.

\begin{figure*}[htb]
    \centering
	\includegraphics[width=0.99\textwidth]{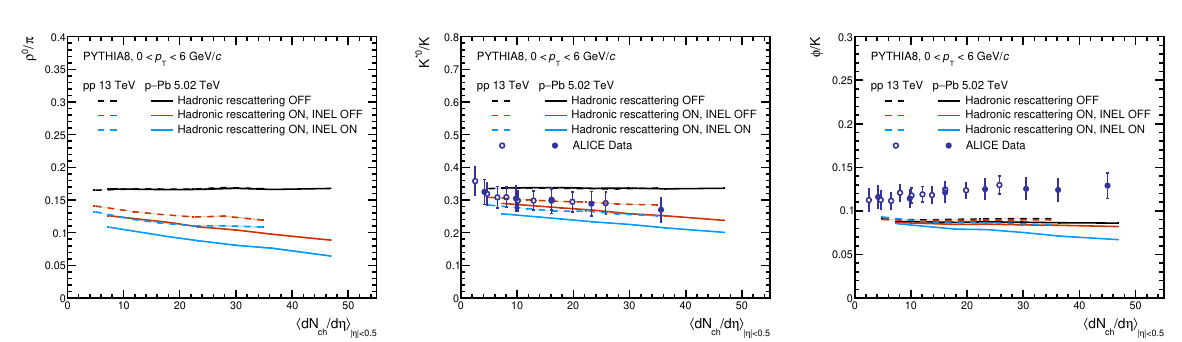} 
    \caption{Particle yield ratios between resonances and stable particles as a function of multiplicity ($\langle \mathrm{d}N_\mathrm{ch}/\mathrm{d}\eta \rangle_{|\eta|<0.5}$) in the range $0<p_\mathrm{T}<6$ GeV/$c$, for pp collisions at $\sqrt{s}=13$ TeV and p--Pb collisions at $\sqrt{s_{\rm NN}}=5.02$ TeV. For $\mathrm{K^{*}(892)^{0}/K}$ and $\phi(1020)/\pi$, the results are compared with ALICE data~\cite{ALICE:2016sak, ALICE:2019etb}.}
    \label{fig:YR_pPb_mult}
\end{figure*}

%The single ratio is also calculated as a function of multiplicity ($\langle \mathrm{d}N_\mathrm{ch}/\mathrm{d}\eta \rangle_{|\eta|<0.5}$) and compared with the data for $\mathrm{K^{*0}/K}$ and $\phi/\mathrm{K}$ in integrated \pt\ region ($0<p_\mathrm{T}<6$ GeV/$c$). For all three particles and both collision systems, the single ratio seems to be multiplicity independent for hadronic rescattering off, while both rescattering on options show decreasing trend with increasing multiplicity. Hadronic rescattering on with inelastic scattering on option shows more suppressed yield ratio compared to inelastic scattering off option in both pp and p--Pb collisions. The ALICE data seems to favor the model when the rescattering option is on for $\mathrm{K^{*0}/K}$, however the model underestimates the data for $\phi/\mathrm{K}$.

Figure 7 shows the particle ratios in the integrated $p_\mathrm{T}$ region as a function of multiplicity ($\langle \mathrm{d}N_\mathrm{ch}/\mathrm{d}\eta \rangle_{|\eta|<0.5}$). Results from pp collisions at 13 TeV (dashed lines) and p--Pb collisions at 5.02 TeV (solid lines) are presented together and compared with the ALICE measurements. In the PYTHIA simulations with hadronic rescattering off, the particle ratios of all three resonances show no clear multiplicity dependence in either collision system, and the values in pp and p–Pb collisions are consistent with each other.
For the $\rho(770)^0$ and $\mathrm{K^{*}(892)^0}$, when hadronic rescattering is enabled, the ratios decrease with increasing multiplicity. The slope of the decrease appears similar for both inelastic scattering on and off options; however, with inelastic scattering on, the ratios are about 3\% lower than those with inelastic scattering off across the entire multiplicity range. This behavior suggests that inelastic scattering processes reduce resonance yields regardless of multiplicity, with only minor differences observed between pp and p--Pb systems.
In contrast, for the $\phi(1020)$ meson, the hadronic rescattering on with inelastic scattering off configuration shows little difference compared to rescattering off. A slight decreasing trend with multiplicity is visible only when the inelastic scattering option is enabled.
When compared with ALICE data~\cite{ALICE:2016sak, ALICE:2019etb}, the $\mathrm{K^{*}(892)^0}/\mathrm{K}$ ratio shows a similar decreasing trend with multiplicity in PYTHIA when the hadronic rescattering option is enabled. The inelastic-off configuration describes the data more accurately, although the inelastic-on results also remain within the 1$\sigma$ experimental uncertainty. For the $\phi(1020)/\mathrm{K}$ ratio, the model generally predicts larger values than observed in the data, while showing a similar lack of multiplicity dependence. In this case, the ALICE results appear to disfavor the PYTHIA predictions with the inelastic-on option.

\subsection{Ratios between the hadronic rescattering option on and off}

\begin{figure*}[htb]
    \centering
	\includegraphics[width=0.99\textwidth]{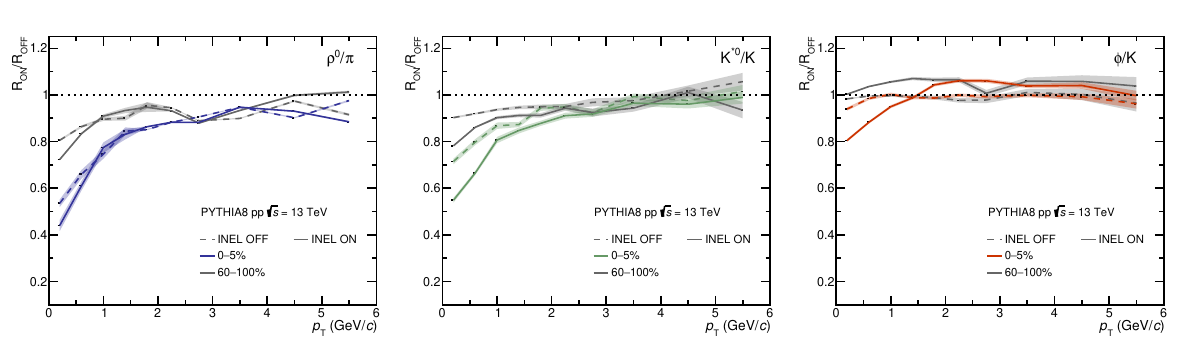}
    \includegraphics[width=0.99\textwidth]
    {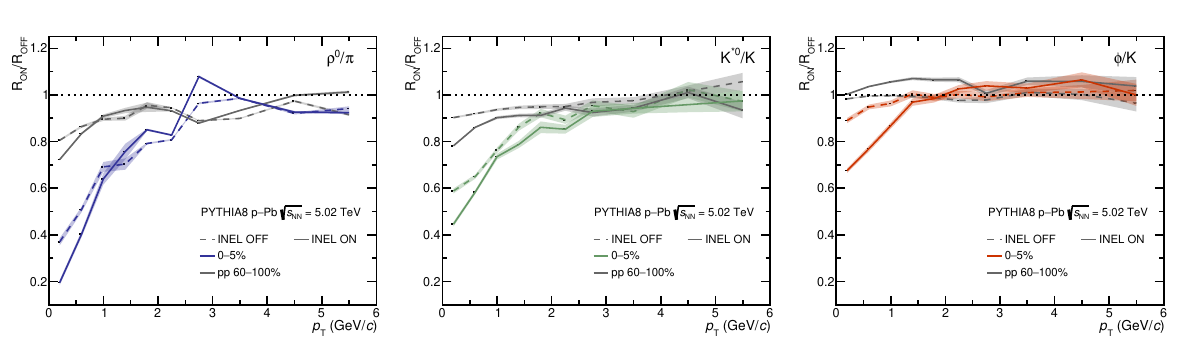}
    \caption{Double ratios of the particle yield ratios between the hadronic rescattering on and off options as a function of $p_\mathrm{T}$ (denoted as $\mathrm{R_{ON}/R_{OFF}}$) in PYTHIA8 for 0--5\% and 60--100\% event classes in pp collisions at $\sqrt{s}=13$ TeV (top) and p--Pb collisions at \snn\ = 5.02 TeV (bottom).}
    \label{fig:DR_pT}
\end{figure*}

%\begin{figure*}[htb]
%    \centering
%	\includegraphics[width=0.99\textwidth]{Figures_4/Dratio_onoff_mult_pp_pPb_all.pdf}
%    \caption{Double ratios of the particle yield ratios between the hadronic rescattering on and off options as a function of multiplicity for $0<p_\mathrm{T}<6$ GeV/$c$ in pp collisions at $\sqrt{s} = 13$ TeV and p--Pb collisions at $\sqrt{s_\mathrm{NN}}=5.02$ TeV.}
%    \label{fig:DR_mult}
%\end{figure*}

\begin{figure*}[htb]
  \centering%
  \includegraphics[width=0.495\textwidth]{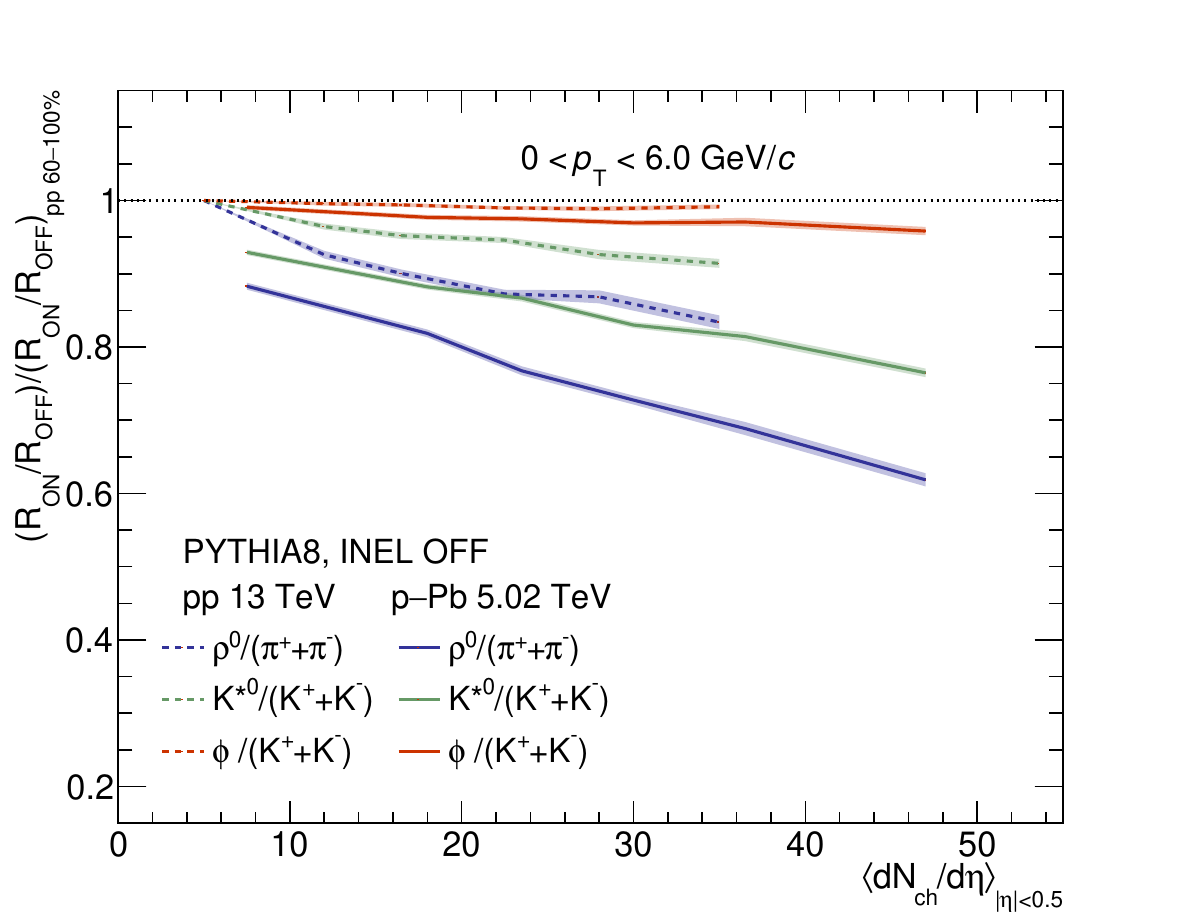} 
    \includegraphics[width=0.495\textwidth]{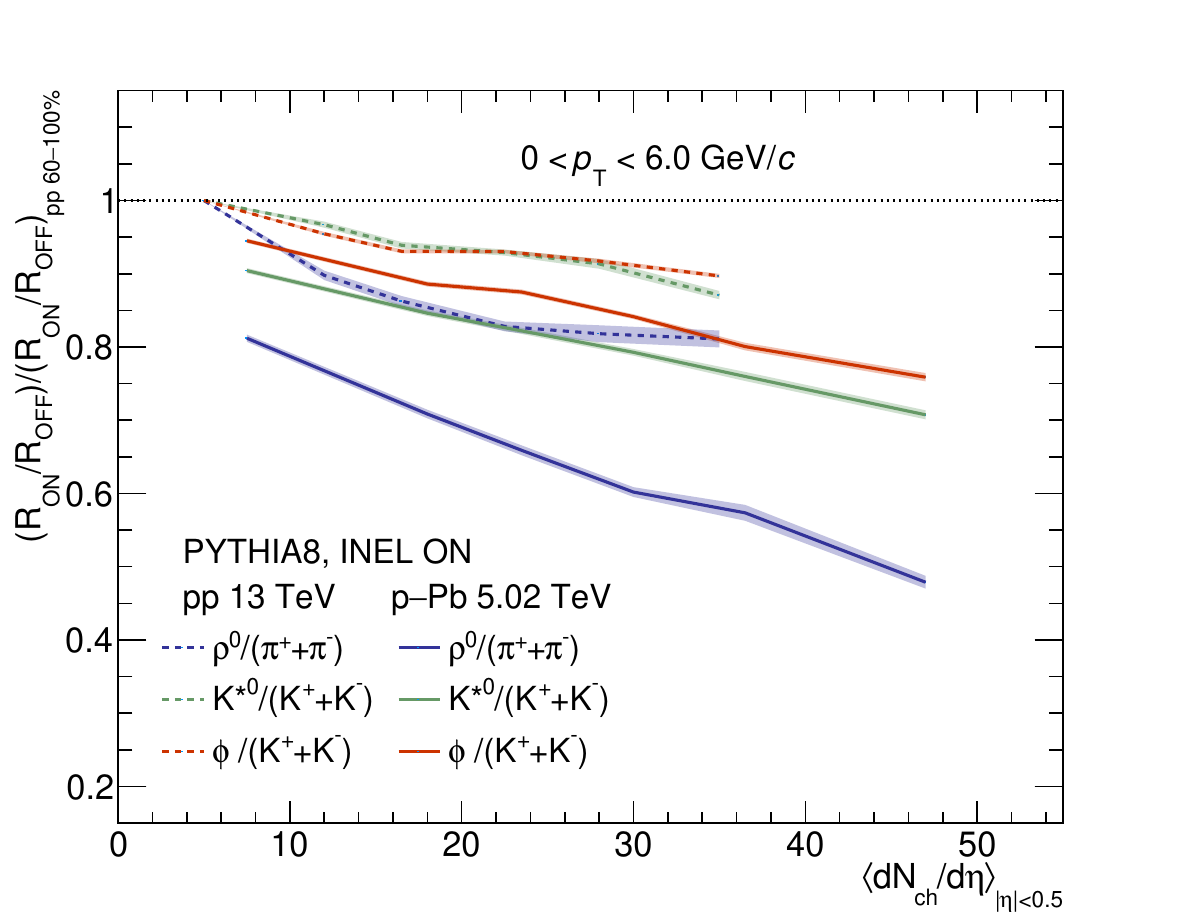}
    \caption{Double ratios of the particle yield ratios between the hadronic rescattering on and off options in 0--5\% events, normalized to that of 60--100\%  (denoted as $\mathrm{(R_{ON}/R_{OFF})_{0-5\%}/(R_{ON}/R_{OFF})_{pp\ 60-100\%}}$) in $0<p_\mathrm{T}<6$ GeV/$c$ from pp collisions at $\sqrt{s}=13$ TeV and p--Pb collisions at $\sqrt{s_\mathrm{NN}}=5.02$ TeV. The left figure represents the double ratio without inelastic option, and the right figure represents with inelastic option.}
    \label{fig:DDR_mult}
\end{figure*}

As the ratios between the particle yield ratios in different multiplicity classes relative to the 60–100\% class in pp collisions without the hadronic rescattering option are already below unity in the low-\pt\ region, as seen in the top panels of Fig.~\ref{fig:YR_pp} and~\ref{fig:YR_pPb}, we isolate the effect of hadronic rescatterings by taking ratios between the particle yield ratios from the rescattering on and off configurations.
The Fig.~\ref{fig:DR_pT} shows the double ratios, denoted as $\mathrm{R_{ON}/R_{OFF}}$, in pp collisions on the top panel and p--Pb collisions on the bottom for inelastic scattering on (solid lines) and off (dashed lines) configurations.
%The double ratios for $\rho(770)^0$ meson shows the largest suppression and then $\mathrm{K^{*}(892)^{0}}$ and $\phi(1020)$ follow.Inelastic scattering on option is more suppressed compared to inelastic scattering off in low-\pt\ region, for both pp and p--Pb collisions.These results highlights that hadronic rescattering effects with inelastic scattering have the most significant impact in the low-\pt\ region, particularly for $\pt<0.8$ GeV/$c$. 

For the $\rho(770)^0$ meson, which has the shortest lifetime, the double ratio in 0--5\% pp and p--Pb collisions exhibits a clear decrease at low $p_\mathrm{T}$. In 60--100\% pp collisions, the double ratio also drops to about 0.7–0.8 for $p_\mathrm{T}<0.5$ GeV/$c$, indicating a noticeable hadronic rescattering effect. When comparing the inelastic on and off options, the results are similar for $p_\mathrm{T}>1$ GeV/$c$, but at $p_\mathrm{T}<1$ GeV/$c$ the inelastic on option shows a lower double ratio, suggesting that additional inelastic scattering further reduces the $\rho(770)^0$ yield.
For the $\mathrm{K^{*}(892)^{0}}/\mathrm{K}$ double ratio ($\mathrm{R_{ON}}/\mathrm{R_{OFF}}$), the overall trend is similar to that of the $\rho(770)^0/\pi$ case. However, reflecting the longer lifetime of the $\mathrm{K^{*}(892)^{0}}$ compared to the $\rho(770)^0$, the decrease in the double ratio is less pronounced.
For the $\phi(1020)/\mathrm{K}$ double ratio, in 0--5\% pp and p--Pb collisions, the trend resembles that of $\rho(770)^0/\pi$ and $\mathrm{K^{*}(892)^{0}}/\mathrm{K}$, with hadronic rescattering reducing the $\phi(1020)$ yield at $p_\mathrm{T}<1.5$ GeV/$c$. The suppression effect is stronger when the inelastic option is on than when it is off. This suggests that the reduction in $\phi(1020)$ yield, despite its long lifetime, is more likely due to scattering of the $\phi(1020)$ itself rather than scattering of the kaons from its decay.
An interesting observation arises in 60--100\% pp collisions: when the inelastic option is off, the double ratio remains consistent with unity, implying negligible hadronic rescattering effects. However, with the inelastic option enabled, the double ratio shows an increase of about 5\% around $p_\mathrm{T}\sim1.5$ GeV/$c$. This indicates that, to quantitatively understand the suppression of resonance yields by hadronic rescattering, it is also necessary to account for the mechanisms of resonance production.

%Also, this double ratios are calculated as a function of multiplicity to study the system-size dependent hadronic rescattering and inelastic scattering effect in Fig.~\ref{DDR_mult}.
%The figure presents that the hadronic rescattering effect is more dominant in p--Pb collisions compared to pp collisions for inelastic scattering on option, and inelastic scattering off option for p--Pb collisions shows similar results with inelastic scattering on option for pp collisions.
%The rescattering seems to affect more as system size becomes larger.

%The results for the 0--5\% and 60--100\% multiplicity classes of pp collisions are shown in the left panel of Fig.~\ref{fig:DR_pT}. In the low-\pt\ region, the double ratios for high-multiplicity events are approximately 20\% lower than those for low-multiplicity events. For p--Pb collisions, presented in the right panel, the suppression is even more pronounced—ranging from 30\% to 40\%—depending on the particle species. These results highlight that hadronic rescattering effects have the most significant impact in the low-$p_{\rm T}$ region, particularly for $\pt<0.8$ GeV/$c$.

To explore the system-size dependence of rescattering effects, double ratios of the particle yield ratios with the hadronic rescattering option on to those with the hadronic rescattering option off are calculated and normalized to the 60--100\% multiplicity class of pp collisions. 
The left panel of Fig.~\ref{fig:DDR_mult} presents these double ratios, indicated as $(\mathrm{R_{ON}/R_{OFF})/(R_{ON}/R_{OFF})_{pp\ 60-100\%}}$, as a function of multiplicity for $0 < p_\mathrm{T} < 6.0$ GeV/$c$ with inelastic scattering off option, and the right panel presents the double ratios with inelastic scattering on option.
The double ratios of inelastic scattering on option decrease more steeply as multiplicity increases compared to the inelastic scattering off option for all three resonances.
Also, within same multiplicity intervals, p--Pb collisions seems to undergo stronger suppression compared to pp collisions.
In inelastic scattering off option, the double ratios are separated with the similar degree, however with inelastic scattering on option, $\mathrm{K^*(892)^0}$ and $\phi(1020)$ show comparable behavior, whereas $\rho(770)^0$ meson exhibits a steeper decline.
Regardless of the configuration, a clear lifetime-dependent suppression hierarchy emerges: the short-lived $\rho(770)^0$ meson shows the strongest suppression, while the long-lived $\phi(1020)$ meson is least affected. 
This indicates that the lifetime of resonances plays a crucial role in their susceptibility to rescattering in the hadronic phase.

%A gradual decrease in the double ratios with increasing multiplicity is observed, with no significant variation among the different particle species. Furthermore, the double ratios appear similar between pp and p--Pb collisions within the same multiplicity intervals.
%In contrast, the right panel of Fig.~\ref{fig:DR_mult} focuses on the low-$p_{\rm T}$ region ($0 < p_\mathrm{T} < 0.8$ GeV/$c$), where rescattering effects are expected to dominate. Here, the double ratios decrease more steeply as multiplicity increases. While the overall values remain similar between pp and p--Pb collisions, the trends differ by particle species. 
%In pp collisions (dashed lines), the double ratios of $\mathrm{K^{*}(892)^{0}/(K^{+}+K^{-})}$ and $\phi(1020)/\mathrm{(K^{+}+K^{-})}$ show comparable behavior, whereas $\rho(770)^{0}/(\pi^{+}+\pi^{-})$ exhibits a steeper decline, reaching as low as 0.7. In p--Pb collisions (solid lines), a clear lifetime-dependent suppression hierarchy emerges: the short-lived $\rho(770)^0$ meson shows the strongest suppression, while the long-lived $\phi(1020)$ meson is least affected. This indicates that the lifetime of resonances plays a crucial role in their susceptibility to rescattering in the hadronic phase.

\subsection{Extraction of the hadronic phase lifetime}

\begin{figure*}[htb]
    \centering
	\includegraphics[width=0.99\textwidth]{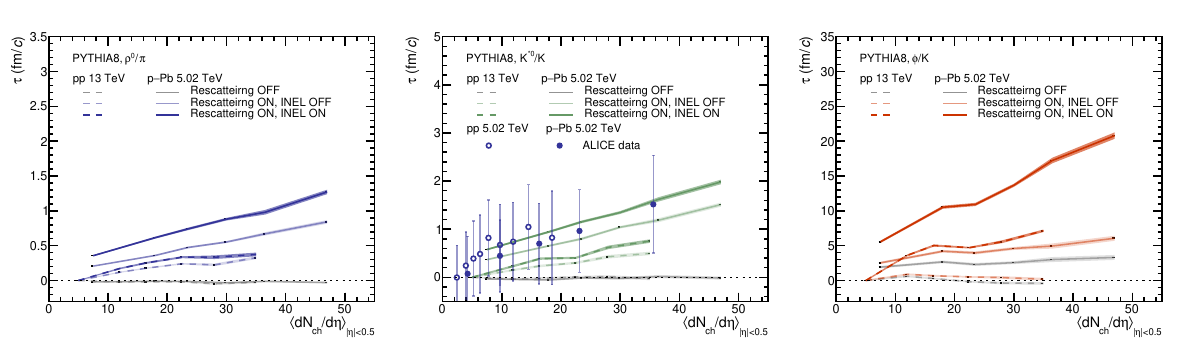} 
    \caption{The lower limit of the lifetime of the hadronic phase extracted using $\rho(770)^0$, $\mathrm{K^{*}(892)^{0}}$ and $\phi(1020)$ mesons as a function of $\langle \mathrm{dN_{ch}/d\eta} \rangle_\mathrm{|\eta|<0.5}$ with and without the hadronic rescattering option, both for inelastic on and off, in PYTHIA8 for pp collisions at $\sqrt{s}=13$ TeV and p--Pb collisions at $\sqrt{s_{\rm NN}}=5.02$ TeV. The particle yield ratios in $0<p_{\rm T}<6$ GeV/$c$ are used for the calculation. For $\tau$ obtained using $\mathrm{K^{*0}/K}$ is compared with the ALICE data~\cite{ALICE:2023edr}.}
    \label{fig:HPTau1}
\end{figure*}

The decreasing trend of the double ratios seen in Fig.~\ref{fig:DDR_mult} suggests that the suppression effect plays a dominant role compared to the regeneration effect due to hadronic rescatterings at the hadronic phase. Assuming that all resonances whose daughter particles undergo hadronic rescatterings are effectively lost, and that regeneration is negligible, the lower limit of the hadronic phase lifetime ($\tau$) can be estimated using the rate function formalism~\cite{ALICE:2019xyr}:
\begin{align}
\left[\mathrm{R_{yield}} \right]_{\mathrm{kinetic}} = \left[\mathrm{R_{yield}} \right]_{\mathrm{chemical}} \times e^{-\tau/\tau_\mathrm{res}},
\end{align}
\label{eq:HPtau}

where $\tau_\mathrm{res}$ denotes the lifetime of the resonance.
As previously mentioned, $\tau_\mathrm{res}$ for $\rho(770)^0$, $\mathrm{K^{*}(892)^{0}}$, and $\phi(1020)$ mesons are 1.3 fm/$c$, 4.16 fm/$c$, and 46.3 fm/$c$, respectively.
The lifetime for a specific \pt\ range of resonance is corrected by the Lorentz factor $\sqrt{1+(\langle p_\mathrm{T} \rangle/mc)^2}$.
Here, $\langle p_\mathrm{T} \rangle$ represents the mean transverse momentum and $m$ is the resonance mass. 
The yield ratios measured in 60--100\% multiplicity events from pp collisions are used as proxies for $\left[\mathrm{R_{yield}} \right]_\mathrm{chemical}$ for other multiplicity classes in pp and p--Pb collisions. Using these assumptions, the lower limit of the hadronic phase lifetime $\tau$ can be extracted.

Fig.\ref{fig:HPTau1} shows the estimated lower limits of the hadronic phase lifetime as a function of charged-particle multiplicity in pp and p--Pb collisions, obtained with and without the hadronic rescattering option in PYTHIA8. The estimates are restricted to the interval $0 < p_\mathrm{T} < 6.0$ GeV/$c$. With hadronic rescattering enabled, $\tau$ increases with multiplicity for both inelastic scattering on and off, in both collision systems, although the magnitude depends on the particle species. 
Notably, within the same multiplicity intervals, the lifetimes extracted from p--Pb collisions are consistently larger than those from pp collisions, consistent with the double-ratio behavior in Fig.\ref{fig:DDR_mult}. 
This suggests that the hadronic phase in p--Pb collisions could be intrinsically longer-lived, likely reflecting the larger system size and higher medium density. Furthermore, $\tau$ values from $\mathrm{K^{*0}/K}$ are compared with ALICE results in pp collisions at $\sqrt{s}=5.02$ TeV and in p--Pb collisions at $\sqrt{s_\mathrm{NN}}=5.02$~\cite{ALICE:2019xyr, ALICE:2023edr}. The data appear to follow the trend of the rescattering-on options in PYTHIA8 for p--Pb collisions, irrespective of the collision system, and exhibit a single multiplicity-dependent trend, unlike the PYTHIA predictions. 
However, due to the large experimental uncertainties, it remains difficult to determine whether the data favor the inelastic on or off configuration. In contrast, when hadronic rescattering is switched off, $\tau$ stays nearly constant and close to zero in both pp and p--Pb collisions, indicating the absence of a hadronic phase in this scenario.

%Since the particle yield ratios in high-multiplicity class is lower than that in low multiplicity class, even without the hadronic rescattering option, we isolate the net rescattering contribution by subtracting the lower limit of the hadronic phase lifetime obtained in the rescattering off case from that of the rescattering on cases. Fig.~\ref{fig:HPTau3} displays the resulting lower limit of net lifetime due to rescattering with and without inelastic scattering as a function of multiplicity for both systems. 
%The lower limit of net lifetime increases with multiplicity for all resonances, but the magnitude varies by particle species. Notably, in the same multiplicity intervals, the lower limit of the lifetimes extracted from p--Pb collisions are consistently larger than those from pp collisions, likewise their double ratio appearing different.
%This suggests that the hadronic phase in p--Pb collisions is intrinsically longer-lived, likely due to the larger system size and increased medium density.

\section {Conclusions}
\label{sec:summary}
This study investigated how the interactions during the hadronic phase affect on the particle yield ratios between short-lived resonances and their stable particles with PYTHIA8 event generator for pp collisions at $\sqrt{s}=13$ TeV and p--Pb collisions at $\sqrt{s_\mathrm{NN}}=5.02$ TeV.
We selected three resonances, $\rho(770)^0$, $\mathrm{K^{*}(892)^0}$, and $\phi(1020)$, and studied the hadronic rescattering effect by comparing the hadronic rescattering options off and hadronic rescattering on with inelastic scattering on and off, as implemented in PYTHIA8.
The results of particle yield ratios as a function of \pt\ showed that the rescattering off option also exhibited a decreasing trend, especially in the low-\pt\ region.
However, for the particle yield ratio as a function of multiplicity for hadronic rescattering off was independent of the multiplicity, unlike the rescattering on options showed decreasing trend.
To isolate the hadronic rescattering effect, the double ratios, particle yield ratios with the rescattering on-to-off option, were measured for 0--5\% high-multiplicity and 60--100\% low multiplicity from pp collisions.
The double ratios showed suppression in the overall \pt\ region, but particularly in the low-\pt\ region, the hadronic rescattering effect seemed to play a dominant role on the ratios.
To obtain the system size dependent particle yield ratio, we measured the $\langle \mathrm{dN_{ch}/d\eta} \rangle_{|\eta|<0.5}$ dependent particle yield ratios for integrated \pt\ region ($0<p_\mathrm{T}<6.0$ GeV/$c$).
The double ratios for the integrated \pt\ region showed significant suppression as observed in the single ratios, as well as showing difference between pp and p--Pb collisions.

The lower limit of the hadronic phase lifetime was further estimated using the yield ratios in 0--5\%, referencing 60--100\% multiplicity class from pp collisions.
%The difference between the rescattering on and off options was used to quantify the contribution from hadronic rescattering.
%The lifetimes of the hadronic phase for both rescattering on and off options were calculated using the particle yield ratio in high-multiplicity 0--5\%, referencing the low multiplicity 60–100\% from pp collisions.
%Then, the lifetime from rescattering off was subtracted from the rescattering on option.
The lower limit of the lifetime showed higher value for p--Pb collisions compared to that of pp collisions.
The observed discrepancy in the lower limit of the hadronic phase lifetimes between pp and p--Pb collisions indicates that the system size plays a key role in determining the duration and strength of final-state interactions.\\

\section*{Acknowledgments}
%\acknowledgments
%\textit{Acknowledgments.---} 
S. Ji is supported by the National Research Foundation of Korea (NRF) grant funded by the Korea government (MSIT) under Contract No. RS-2024-00414492 and 2024 BK21 FOUR Graduate School Innovation Support funded by Pusan National University (PNU-Fellowship program).
S. Ji and S. Lim are supported by the National Research Foundation of Korea (NRF) grant funded by the Korea government (MSIT) under Contract No. RS-2008-NR007226. 
We also acknowledge technical support from KIAF administrators at KISTI.

\nocite{*}
\bibliographystyle{apsrev4-1} 
\bibliography{paper}% Produces the bibliography via BibTeX.

\end{document}

%% file: def.tex
\newcommand{\s}            {\ensuremath{\sqrt{s}}}
\newcommand{\pt}           {\ensuremath{p_\mathrm{T}}}

\newcommand{\snn}          {\ensuremath{\sqrt{s_{\rm NN}}}}